\documentclass{aa}
\usepackage{graphicx}
\usepackage{amssymb}
\usepackage{amsmath}
\usepackage{natbib}

\def\ds{{$\delta$~Scuti~}}
\def\dss{{$\delta$~Scuti~stars~}}
\def\Hip{{\emph{Hipparcos~}}}
\def\mv{{M_{\mathrm{v}}}}
\def\mmv{{(m-M)_{\mathrm{v}}}}
\def\bv{{B_2-V_1}}
\def\teff{{T_{\mathrm{eff}}}}
\def\aosc{{A_{\mathrm{osc}}}}
\def\vsini{{\vr\!\sin\!i}}
\def\vr{{V}}
\def\vrp{{V}^{\prime}}
\def\aper{{$\alpha$-Persei}}
\def\rs{{r_{\mathrm{s}}}}
\def\sigrs{{\sigma(\rs)}}
\def\sigr{{\sigma(r)}}
\def\mmsun{{M/M_{\sun}}}

\def\o {{\mathcal{O}}}
\def\dst{\displaystyle}
\def\ip {{$i^{\prime}$}}
\def\op {{\omega^{\prime}}}

\begin{document}

   \title{A study of correlation between the oscillation amplitude
and stellar parameters for \dss in open clusters}

   \subtitle{Toward selection rules for \dss oscillations.}

   \titlerunning{A study of correlation for \dss in open clusters}
   \authorrunning{Su\'arez et al.}

   \author{J-C Su\'arez
          \inst{1}, E. Michel \inst{1}, F. P\'erez Hern\'andez \inst{2,3},
          Y. Lebreton \inst{1},
          Z.P. Li \inst{4} \and L. Fox Machado \inst{2}}

   \offprints{J-C Su\'arez}

   \institute{LESIA, Observatoire de Paris-Meudon, FRE 2461 \\
              \email{JuanCarlos.Suarez@obspm.fr},
              \email{Eric.Michel@obspm.fr},
          \email{Yveline.Lebreton@obspm.fr}
         \and
             Instituto de Astrof\'{\i}sica de Canarias (IAC), E-38200 La Laguna, Tenerife, Spain \\
             \email{fph@ll.iac.es},
         \email{lfox@ll.iac.es}
     \and
         Departamento de Astrof\'{\i}sica, Universidad de La Laguna, Tenerife, Spain
         \and
             National Astronomical Observatories, Chinese Academy of
             Sciences, C-100012 Beijing, China \\
         \email{lizhi@bao.ac.cn}}

   \date{Received 6 March 2002 / Accepted 3 April 2002}

   \abstract{In the present work, we study correlations between stellar
fundamental parameters and the oscillation amplitude for \dss. We
present this study for a sample of 17 selected \dss belonging to 5
young open clusters. Taking advantage of properties of \dss in
clusters, we correct the photometric parameters of our objects for
the effect of fast rotation. We confirm the benefit of applying
such corrections in this kind of studies. In addition, the
technique used for this correction allows us to obtain an
estimation of stellar parameters like the angle of inclination and
the rotation rate, usually not accessible. A significant
correlation between the parameter \ip (estimation of the angle
of inclination of the star) and
the oscillation amplitude is found. A discussion and
interpretation of these a priori surprising results is proposed,
in terms of a possible selection rule for oscillation modes of
\dss. \keywords{Stars:~$\delta$~Sct -- Stars:~rotation --
Stars:~statistics -- Stars:~oscillations -- Stars:~fundamental
parameters -- Galaxies:~star clusters}}

\maketitle
%

\section{Introduction}

 The \dss are variables located in the
lower part of the Cepheid instability strip with spectral types
from A2 to F0. These pulsators, when belonging to the Main Sequence,
 seem particularly suitable for
determining the extent of the convective core and internal
rotation rate, and thereby for probing poorly understood
hydrodynamical processes occurring in deep stellar interiors.
Within the last decade great efforts have been made in developing
the seismology of \dss \citep{Breger00,Handler00}.
However, several aspects of the pulsating behavior of these stars
(e.g. the number and the determination of the excited modes)
within the instability strip are not completely understood.

Several statistical works investigated potential correlations
between the oscillation amplitude and different stellar
parameters. Particularly, a correlation between the amplitude,
period and absolute magnitude of low amplitude \dss has been
established \citep{Anto81}. In a preliminary work for \dss in the
Praesepe cluster, \citet{Sua01} have suggested
the importance of taking into account the effect of fast rotation
as proposed by \citet{MiHer99} in such
studies.

\begin{table}
  \begin{center}
    \caption{Main characteristics of the 5 open clusters used in this
work. Different columns represent the cluster name, distance
modulus, metallicity (columns 3 and 4), reddening and age.} \vspace{1em}
    \renewcommand{\arraystretch}{1.1}
    \begin{tabular}[h]{lccccc}
      \hline\hline
       Cluster & $\mmv$ & $[M/H]$  & $Z$ & $E(B_2-V_1)$ & Age  \\
            & (mag)  &        &     & (mag) & (Myr)\\
      \hline
          Praesepe & 6.28 & 0.170   & 0.025 & 0.0  & 650 \\
          Pleiades & 5.50 & -0.112  & 0.014 & 0.05 & 130 \\
            Hyades & 3.33 & 0.143   & 0.012 & 0.0  & 600 \\
          Coma Ber & 4.70 & -0.048  & 0.016 & 0.0  & 430 \\
             \aper & 6.23 & -0.05   & 0.016 & 0.09 & 90 \\
      \hline\hline
      \end{tabular}
    \label{tab-amas}
  \end{center}
\end{table}

Here we extend this work to 5 open clusters, in order to estimate
the impact of the mentioned technique in the studies of
correlation. In addition, these tools allow us to obtain a rough
estimation of several stellar parameters like the angle of
inclination of the star or the rotation rate, which are not
usually accessible. We present a study of single correlation
between the parameters mentioned herebefore and the oscillation
amplitude. For values indicating a reasonable degree of
correlation, we also search for a possible functional dependence.

 This paper is organized as follows: in Section~2 we present
the observational material and the main characteristics of the
selected open clusters. Details on the computation of models are
given in Section~3. In Section~4, we introduce the stellar
parameters corrected for the effect of fast rotation. Section~5 is
dedicated to describe the technique of correlation and the
analysis of the results. Finally, we bring up our conclusions in
section 6.


\section{Observational data}
The oscillation amplitude (hereafter $\aosc$), as classically, is taken
to be the highest peak-to-peak measurement in the observed light
curve. Values for $\aosc$ used here are taken from \citet{LiMi99} and
 references therein. The interpretation of this parameter is not
straightforward; it can be considered as a maximum constructive
interference of the observed modes. Thus, in such an
interpretation, and for a given angle of inclination, it can be taken as an
indicator of the energy of these modes.With such an estimation, it is
 evident that short time series
might induce an underestimation of the amplitude, especially in
presence of beating phenomena due to close frequencies. In order
to take this effect into account, we associate with each amplitude
estimation a weight corresponding to the length of the time
series. In Table~\ref{tab-phot} are listed such values where
number 1 corresponds to effective time series shorter than 4
hours; 2, between 4 and 20 hours; 3, between 20 and 200 hours, and
finally 4 corresponds to time series longer than 200 hours.
\begin{table}
  \begin{center}
    \caption{Indivual parallaxes for selected \dss of Hyades. $\pi$ represents the
    parallax in miliarcseconds and $\mmv$ is the resulting distance modulus.}\vspace{1em}

    \renewcommand{\arraystretch}{1.2}
    \begin{tabular}[h]{cccc}
      \hline
        Star & $\pi$  & $\mmv$ & \\
             & (mas)  & (mag) & \\
      \hline
      \object{HD\,27397} & $22.38\pm0.37$ & $3.25\pm0.05$ \\
      \object{HD\,27459} & $20.87\pm0.32$ & $3.40\pm0.10$ \\
      \object{HD\,27628} & $22.30\pm0.36$ & $3.25\pm0.05$ \\
      \hline
      \end{tabular}
    \label{tab-paral}
  \end{center}
\end{table}
The five young open clusters considered  are: Praesepe, Pleiades,
Hyades, Coma~Ber and \aper, with ages between~$90$ and $700$~Myr.
Detailed information concerning metallicity, age, distance modulus
and reddening are listed in Table~\ref{tab-amas}. Apparent magnitudes
and colour indexes are taken from the Rufener's cataloge \citep{Ruf99}.

 For \aper, the estimation of the global metallicity $[M/H]$ by spectroscopic
measurements is
taken from \citet{Pinso98}. For the other four clusters, we used
metallicities derived by M. Grenon from Geneva photometry of single stars with spectral
type in the range F4-K3 \citep{Gre00,Rob99}. 
To correct the visual absorption due to
interstellar matter, the relation
\begin{center}
$A_{\mathrm{v}} = 3.6\,E(\bv)$
\end{center}
was used. The
intrinsic colours in the Geneva photometric system were obtained
from \citet{Nico81}.

The distance moduli for Coma ($4.70\pm0.04$~mag) and Praesepe ($6.28\pm0.12$~mag) are
derived from mean cluster parallaxes computed using
\Hip~intermediate data \citep{Rob99}. In the case
of Pleiades, the estimation of the distance modulus is still a
subject of discussion. The \Hip values ($5.36\pm0.06$) and the one obtained
by \cite{Pinso98}, $5.60\pm0.06$~mag illustrate the range of values found
in the literature \citep{Stel01}. For the Hyades, we have
also access from literature, to the individual parallaxes of our
target \dss \citep{Bruij01} which are listed in
Table~\ref{tab-paral}. For the other stars we use an estimation of
$3.33\pm0.01$~mag by \citet{Perry98} from \Hip observations. Finally for
\aper, we used
a distance modulus of $6.23\pm0.06$ from \citet{Pinso98}.

  The projected velocity ($\vsini$) values listed in
Table~\ref{tab-phot} are taken from \citet{Uesu82}.


\section{Stellar models and isochrones}

\begin{figure*}[h]
\includegraphics[width=9.5cm]{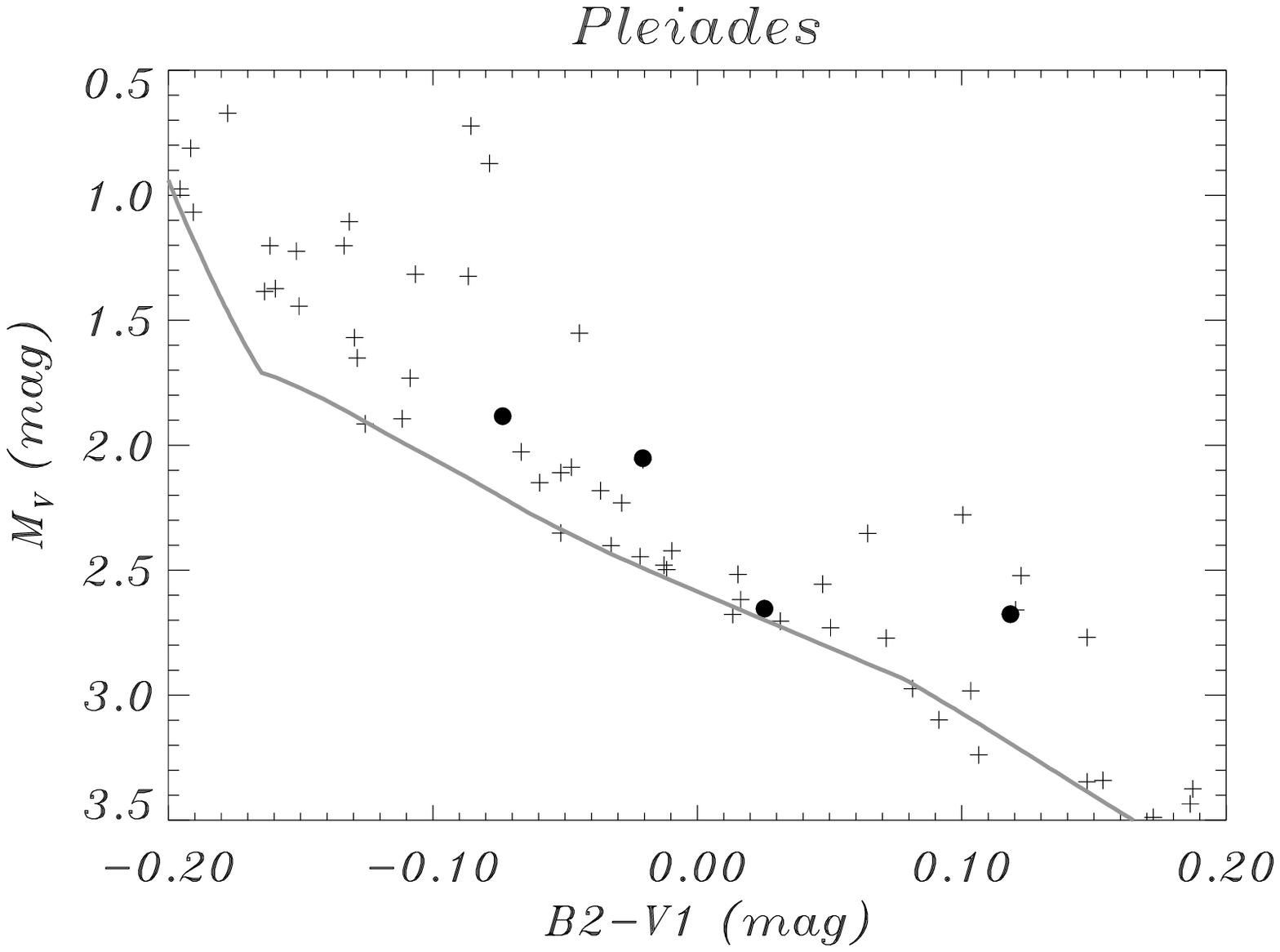}
\includegraphics[width=9.5cm]{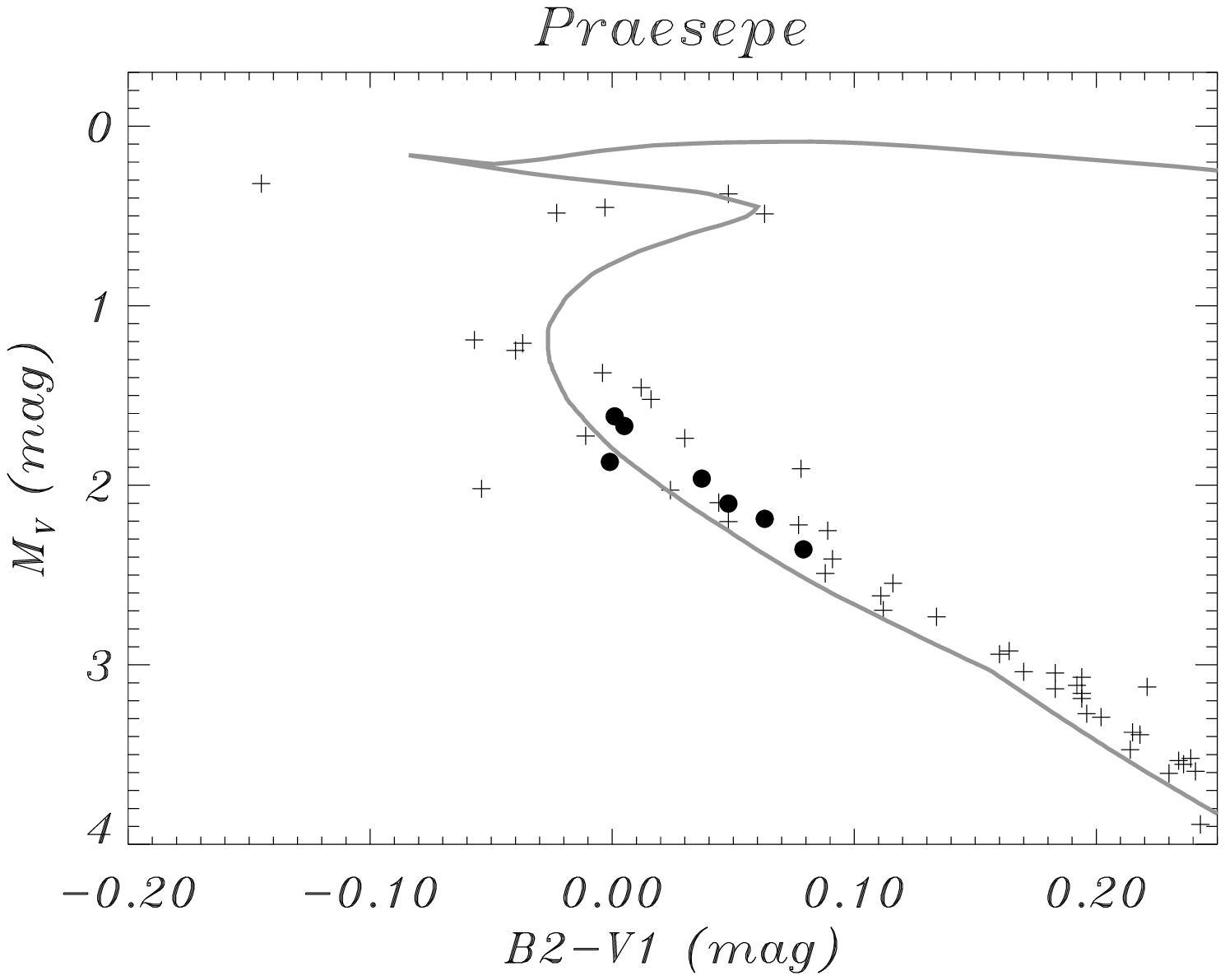}
\includegraphics[width=9.5cm]{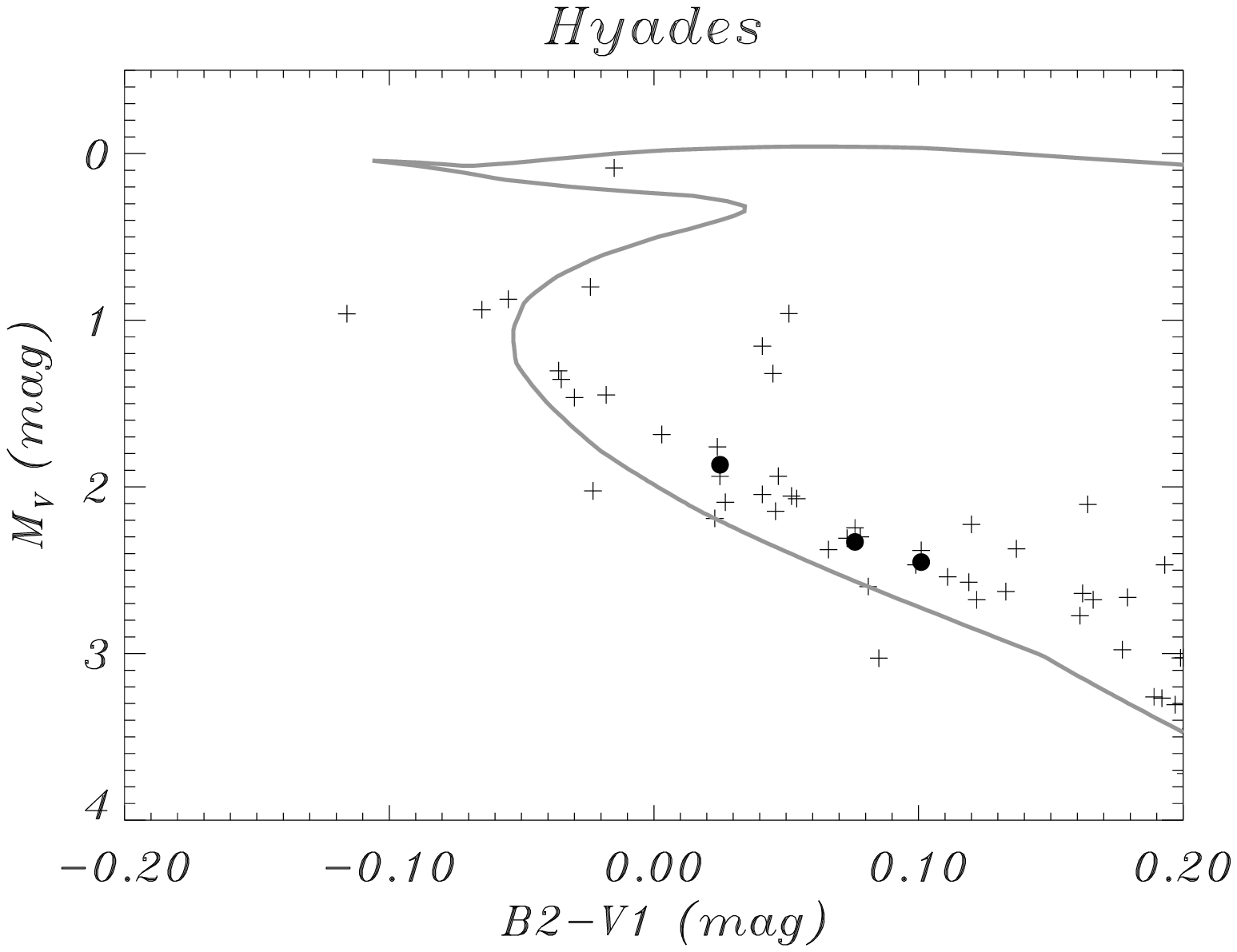}
\includegraphics[width=9.5cm]{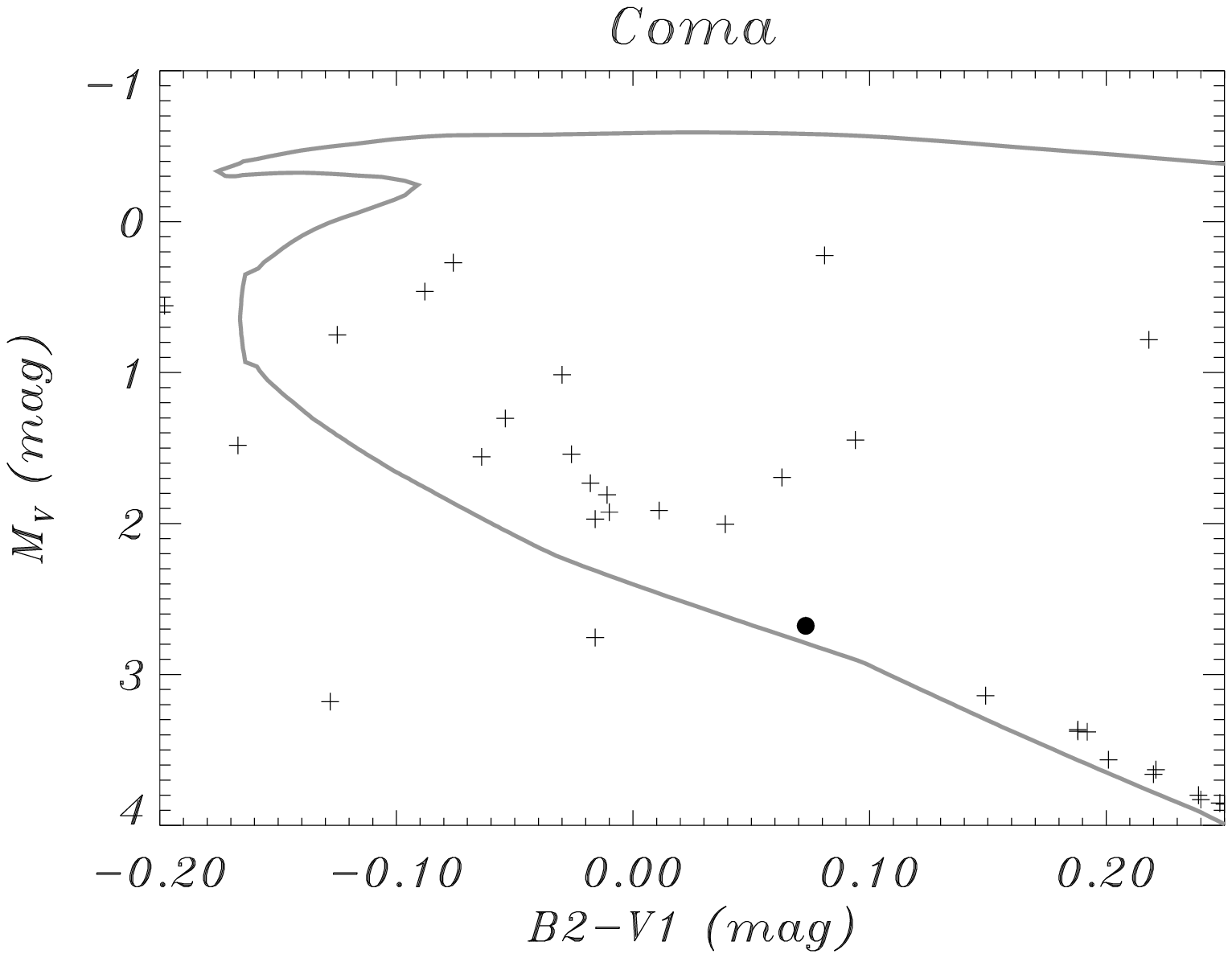}
\includegraphics[width=9.5cm]{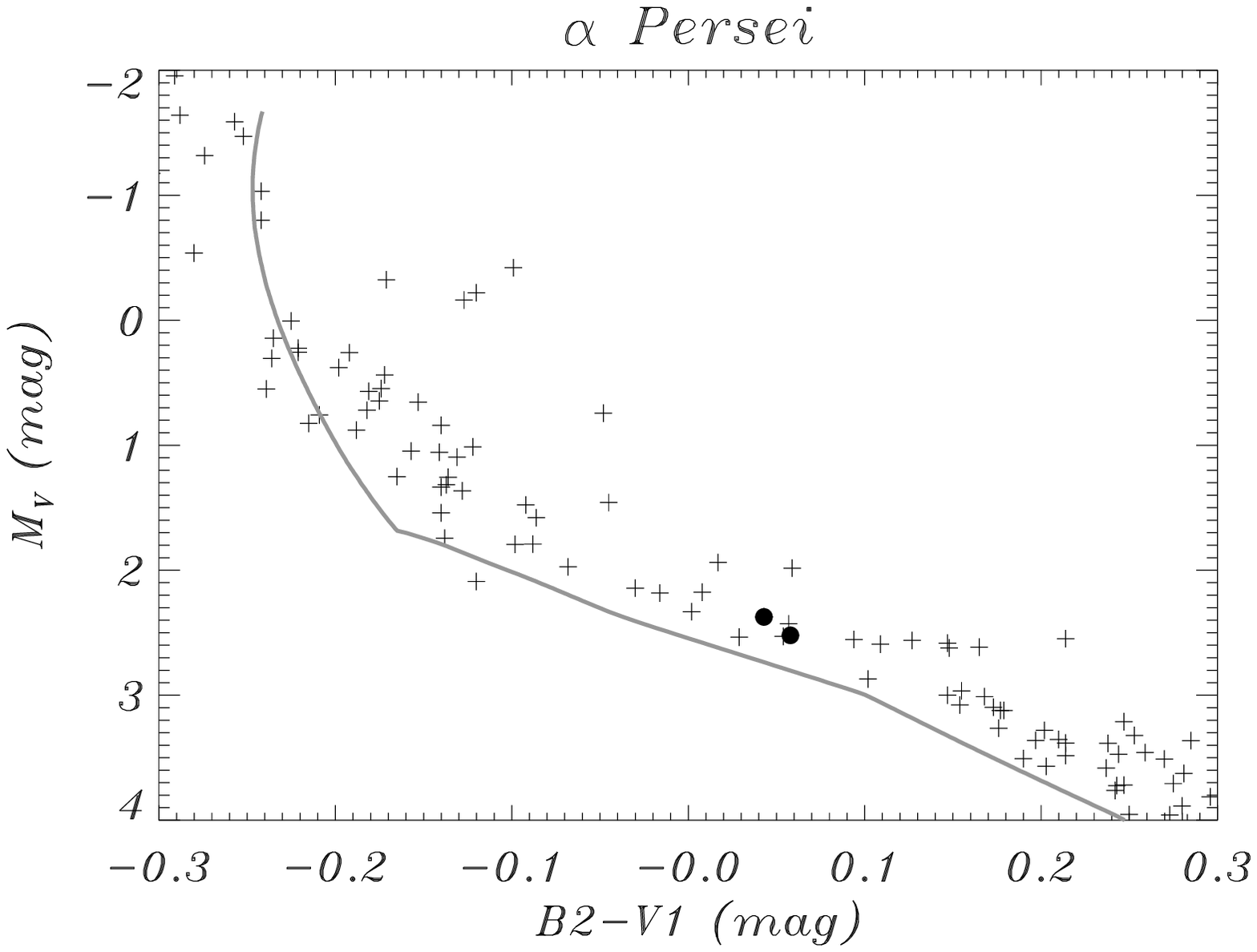}
\caption{Colour -- Magnitude diagrams for the selected 5 open
clusters. Filled circles and crosses represent respectively the
observed \ds and other stars belonging to each cluster. The
correction for distance and reddening is taken into account as
mentioned in the text (see Section~2). }\label{fig-isoc}
\end{figure*}

\begin{figure*}[h]
\includegraphics[width=9cm]{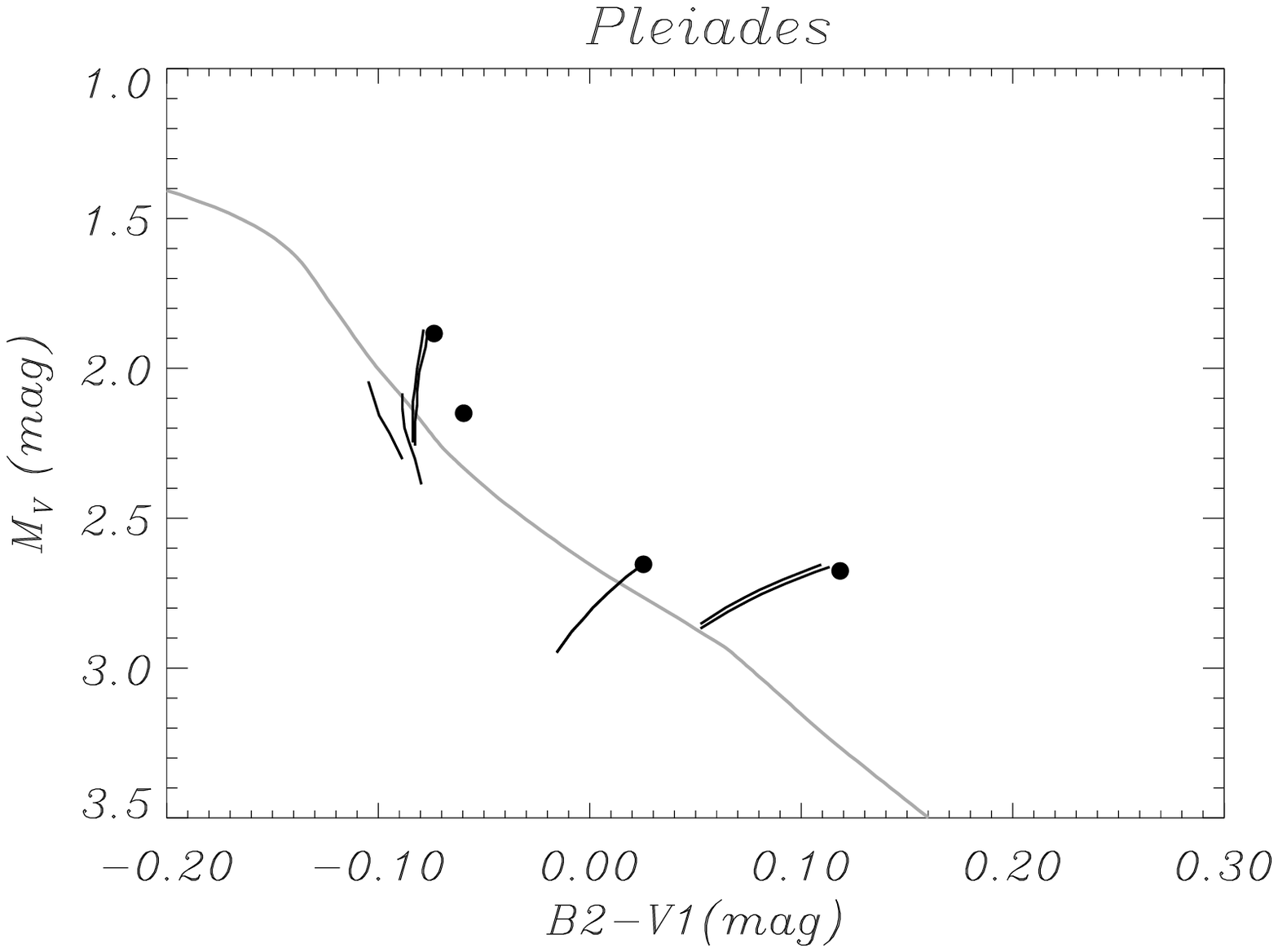}
\includegraphics[width=9cm]{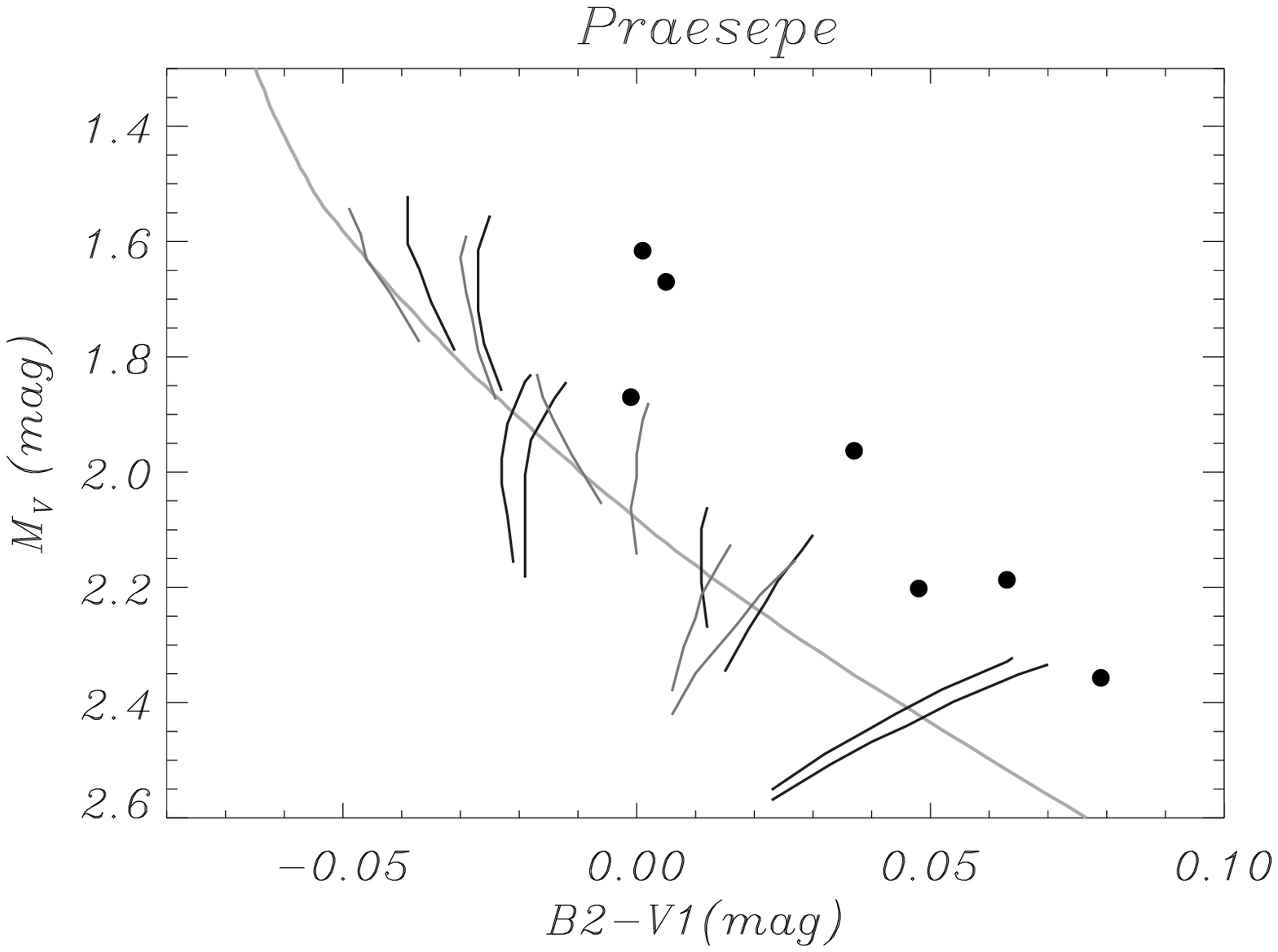}
\includegraphics[width=9cm]{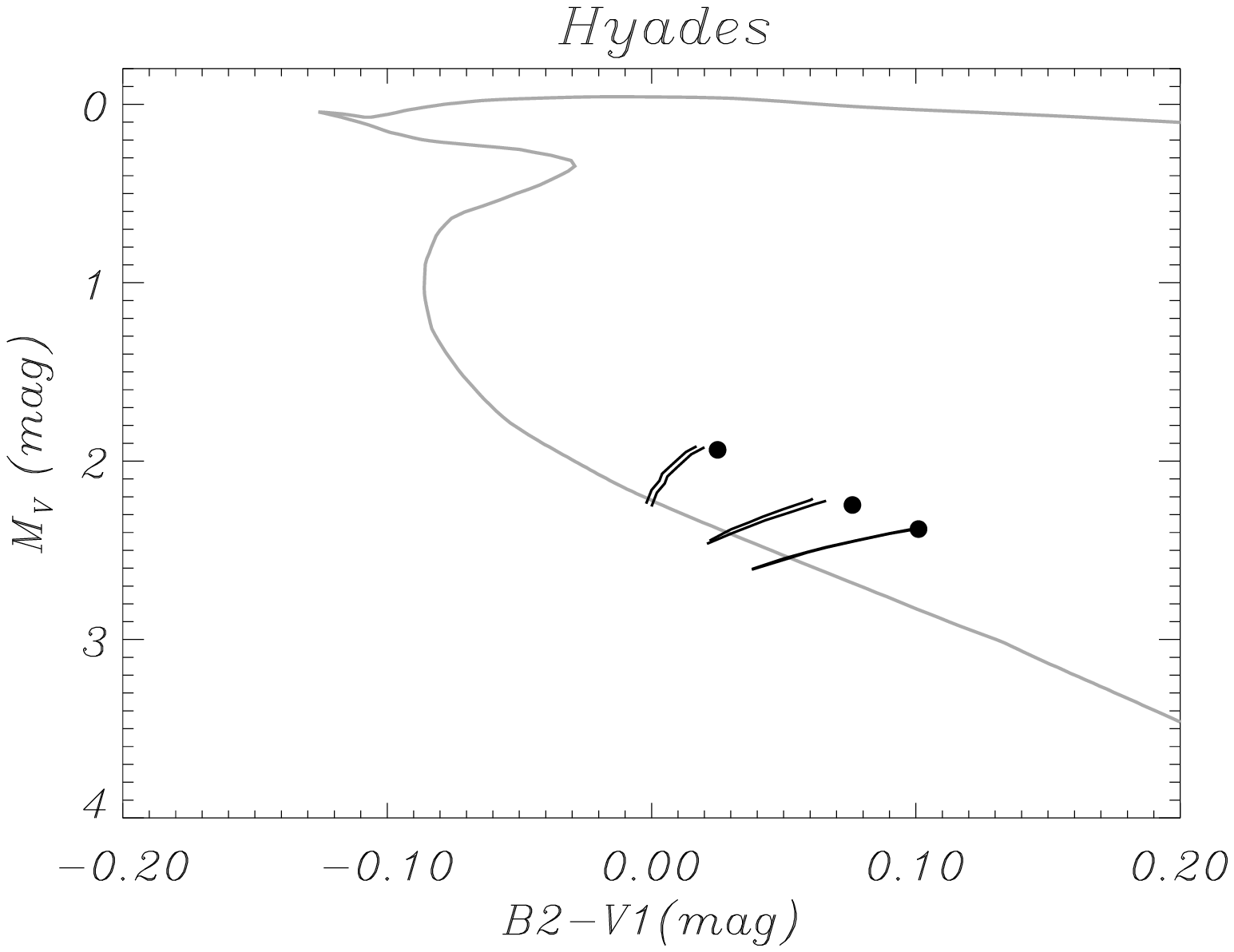}
\includegraphics[width=9cm]{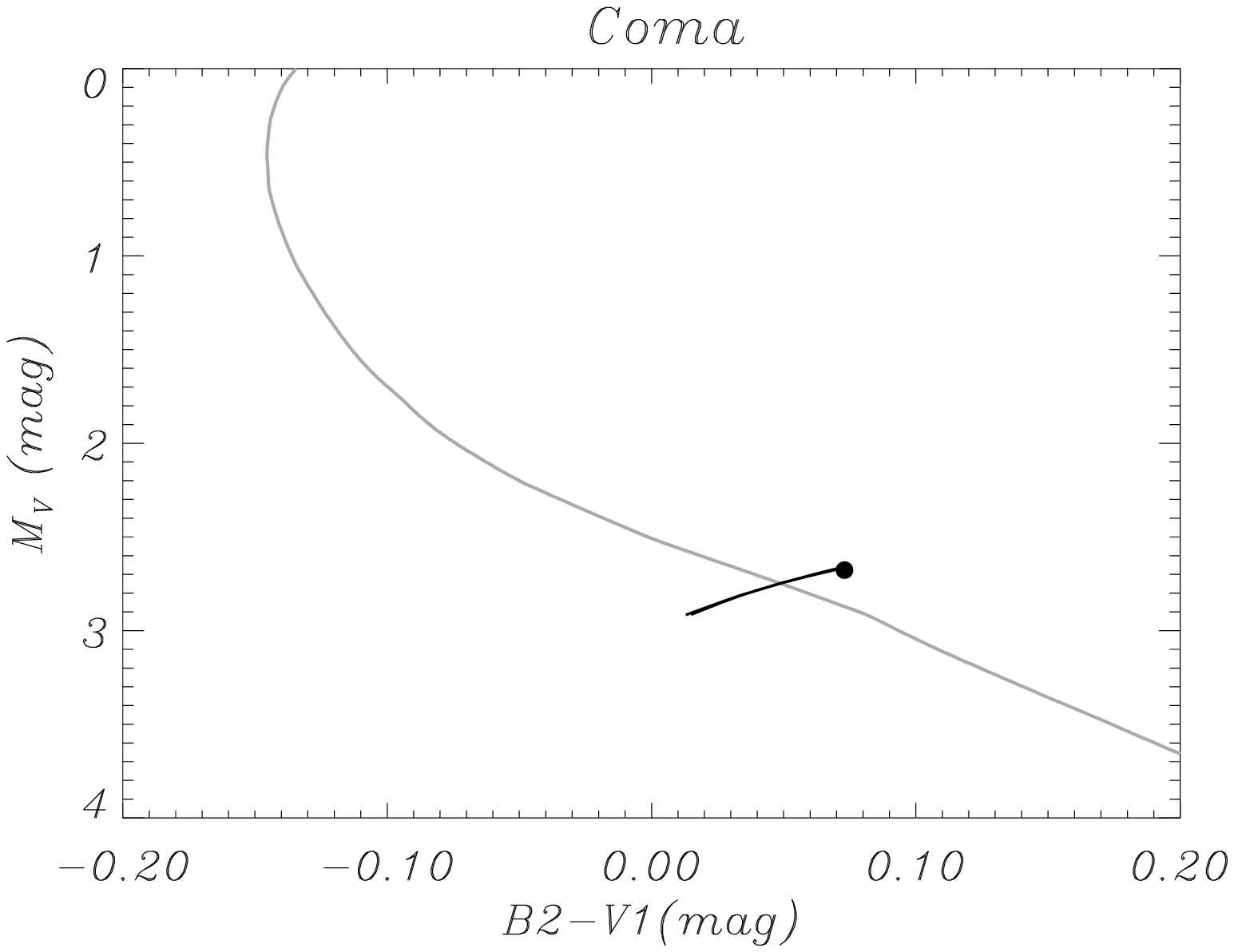}
\includegraphics[width=9cm]{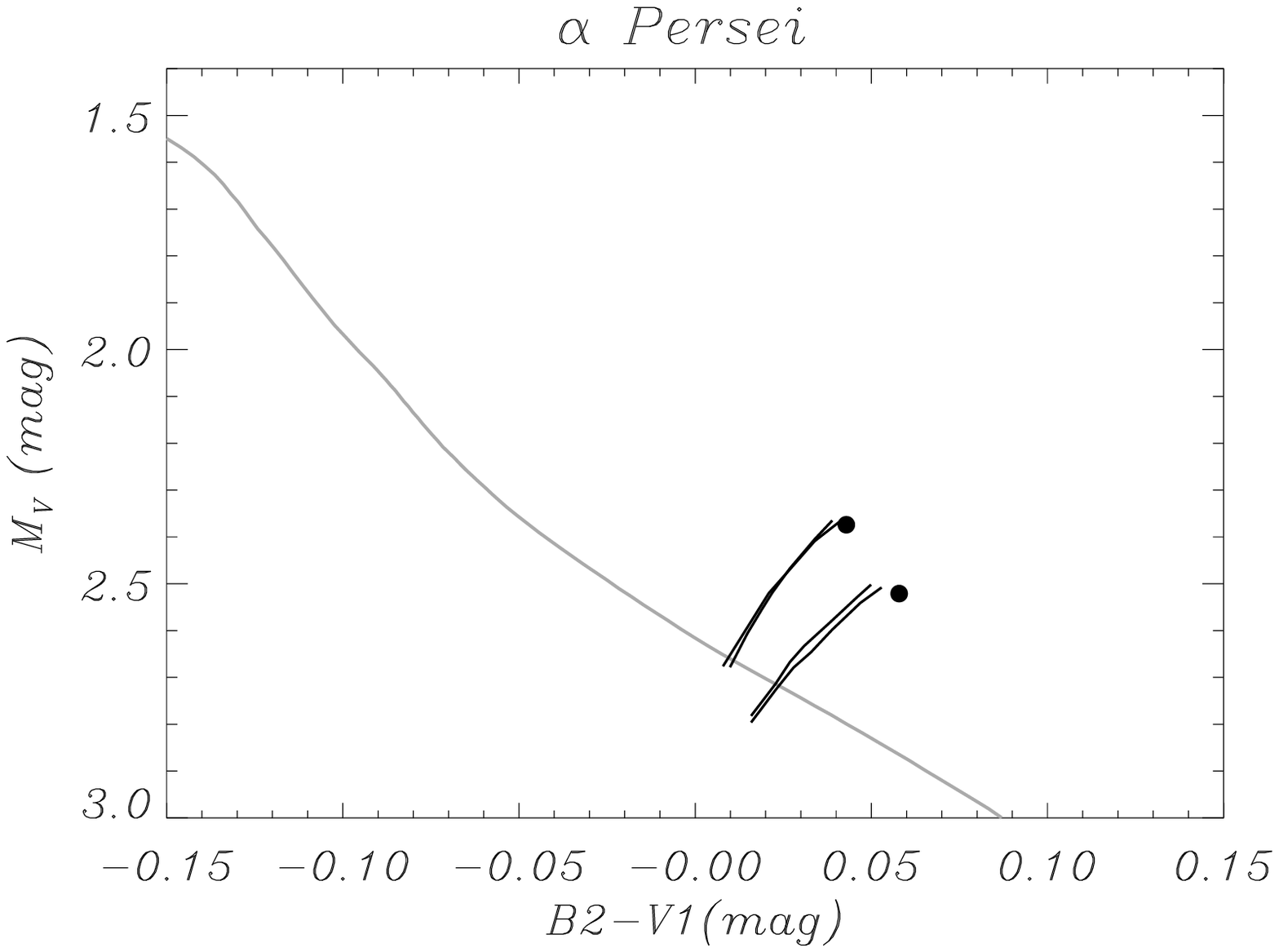}
\caption{Colour -- Magnitude diagrams for the selected 5 open
clusters. Filled circles represent the observed \ds. The
continuous line is the isochrone corresponding to each cluster.
Segments (two per star) represent the possible non-rotating
counterparts for each observed star (see details in Section~4).}
\label{fig-corr}
\end{figure*}

 The stellar evolutive models have been computed using the evolution code
CESAM \citep{Morel97}.
We consider input physics appropriate to the mass range covered by
\dss. Metallicity values ($Z$) used to compute models, are deduced
from $[M/H]$ assuming $Y_{\mathrm{pr}}=0.235$ and
$Z_{\mathrm{pr}}=0$ as helium and heavy elements primordial
concentrations, and an enrichment ratio of $\Delta Y/\Delta Z=2$. We work with
OPAL radiative opacity tables
\citep{Igle96}. For the atmosphere reconstruction, the
Eddington's~$T(\tau)$ law is considered. Convection is described
according to the classical mixing-length theory, with a solar calibrated mixing
length parameter $l_{m}=1.61\mathrm{H}p$, where $\mathrm{H}p$ is the local
pressure scale-height. For all the models, we take into account an
overshooting of the mixed convective core over a distance
$\alpha_{\mathrm{ov}}=0.2\mathrm{H}p$ following the prescription
of \citet{Scha92}.

  Sets of evolutionary sequences were computed for masses between $1 M_{\sun}$ and
$5 M_{\sun}$ from the zero-age Main Sequence to the subgiant
branch. With representative sequences for each cluster, we obtain
the corresponding isochrones using the Geneva isochrone program.
These isochrones were then transformed into a $\mv-(\bv)$ diagram
using the calibration of \citet{Kunzli97} for $(\bv)$ and the
calibration of \citet{Schm82} for $\mv$.

  In Fig.~\ref{fig-isoc} we present colour-magnitude~(hereafter CM)
diagrams with the observed data for each cluster. Continuous lines
correspond to their associated isochrones, given the distances and
ages listed in Table~\ref{tab-amas}. Potential binarity and the
effect of fast rotation discussed in Section~4 are expected to
induce systematic shifts toward higher luminosity and lower
effective temperature, compared to single non-rotating stars.
Thus, in order to compensate both effects, the fit of
isochrones has been made adjusting them to the bottom envelope of
the cluster in the CM diagrams.

In the case of Pleiades, we use our best fit corresponding to a
distance modulus of $5.5$~mag, which is approximately half way 
between the main sequence fitting result and \emph{Hipparcos}
distance (see Section~2). The corresponding
age found is $130$~Myr.

\section{Stellar parameters corrected for the effect of fast
rotation}

We will not enter here in the details of the method, already
described in \citet{Pe99}. Following
this paper, we apply the method to selected \dss for each of
our five clusters. In Fig.~\ref{fig-corr} we illustrate such
corrections for the five clusters. For a given $\vsini$, segments
represent the potential position of the star, in absence of
rotation, on the CM diagram. The position of this non-rotating
copartners varies with the angle of inclination, $i$, and the
rotation rate, ${\omega}$. This last dimensionless quantity is
defined by
\begin{equation}{\omega\equiv\dst\frac{\Omega}{\tilde{\,\Omega}_{c}}~~~~
\text{with}~~~~{\tilde\Omega}^2_{\mathrm{c}}\equiv
\frac{8\,GM}{27{R_{\mathrm{p}}}^3}\nonumber}
\end{equation}
where $\Omega$ is the angular rotational velocity of the star;
${\tilde\Omega_{\mathrm{c}}}$ is the angular rotational velocity
that a star with the same mass ($M$) and polar radius
($R_{\mathrm{p}}$) would have if the centrifugal force balanced
the gravitational attraction at the equator.
 The non-rotating counterparts are thus, computed as a function of
the angle of inclination,
from $i=90^\circ$ ($\omega_{\mathrm{min}}$) to $i_{\mathrm{min}}$
obtained for $\omega_{\mathrm{max}}=0.95$.

 For each selected \ds star we compute two associated
segments corresponding to corrections assuming an estimation of
the error in $\vsini$ of $\pm 10\,\%$. Finally, models at the
intersection of segments with the isochrone will be our
non-rotating counterparts, giving a set of corrected photometric
parameters, like absolute magnitudes ($\mv$) and ($\bv$) colour
indices.
In addition to this, the method gives us the possibility
of obtaining an estimation
of other stellar parameters like the
angle of inclination of the star, the radial velocity, and the
rotation rate, given
in Table~\ref{tab-corr}.

\begin{table}
  \begin{center}
    \caption{Correlation between parameters corrected for rotation and the $\aosc$
    for our sample of 17 \dss. The Spearman Rank order coefficient is represented by $\rs$;
    r is the linear Pearson coefficient; $\sigrs$ $\sigr$ are the errors in
    $\rs$ and $r$ respectively, and p is a probability of deviation
    from null-hypothesis.}
    \vspace{1em}
    \renewcommand{\arraystretch}{1.2}
    \begin{tabular}[h]{cccccc}
      \hline
        Parameter & $\rs$ &  $\sigrs$ & p($\rs$) & r  & $\sigr$ \\
      \hline
     $\mv$   & 0.4319 & 0.1384 & 0.0834 & 0.4569 & 0.1431 \\
     $\bv$   & 0.2725 & 0.0691 & 0.2898 & 0.2484 & 0.0571 \\
     \ip     & 0.7481 & 0.0447 & 0.0005 & 0.6553 & 0.0447 \\
     $\vsini$& 0.7663 & 0.0364 & 0.0003 & 0.7122 & 0.0063 \\
     $\vrp   $& -0.2016 & 0.0360 & 0.4376 & -0.1810 & 0.0290 \\
     $\mmsun$& 0.3895 & 0.1251 & 0.1221 & 0.4253 & 0.1367 \\
     $\op   $& -0.1260 & 0.0140 & 0.629 & -0.1223 & 0.0133 \\
   \hline
      \end{tabular}
    \label{tab-stat}
  \end{center}
\end{table}

\section{Searching for correlations with the oscillation amplitude}

\subsection{The errors}

We search for correlations between, on one side, the $\aosc$
parameter, and on the other side, stellar parameters essentially
obtained after correcting for the effect of fast rotation. Here we
discuss the error on these parameters.

For $\aosc$, we estimate $0.005$ and $0.01$ as minimum and
maximum typical errors and we calculate error bars assuming a
linear relation with the weights listed in Table~\ref{tab-phot}.

Considering the stellar parameters, we have to deal with different
error sources: input photometric indexes, $\vsini$ values,
distances, models and calibrations (isochrones), and of course,
the method to correct parameters for the effect of rotation.

The photometric input parameters, basically the $m_{V}$ and $\bv$,
have typical observational errors around $10^{-3}$ magnitudes.
This is between 10 to 100 times smaller than the rotation effect
found for these parameters \citep{Pe99,Sua01}. The error in the
estimation of $\vsini$ (estimated to 10$\%$) is probably a
dominant factor here. It can induce errors up to 0.15 and 0.02 mag
in $\mv$ and $\bv$ corrected values respectively. The error on the
fit of the isochrones (including those on the distance
determination and on the calibration between $\bv$ and $\teff$)
can be of the same order than the one associated to $\vsini$.
However they are expected to be systematic, at least for objects
within the same cluster.

The errors associated more specifically to the models and the
correction for fast rotation are difficult to estimate, because we
lack elements to compare with. Although for $\mv$ and $\bv$, they are
expected to be systematic to a good extent, for the rest of estimated
stellar parameter we cannot guarantee it. In particular for $i$, $V$ and
$\omega$, we can only pretend to obtain rough estimations, which will
be indicated with a ($^\prime$).

We thus, consider that the error on $\vsini$ is representative of
the non-systematic errors here, being the most important source of
errors when searching for correlations. In addition, other
observational aspects like unknown binarity, differences with the
metallicity of the cluster and possible not detected differential
extinction inside the cluster, can have an influence.

\begin{figure*}
\includegraphics[width=9cm]{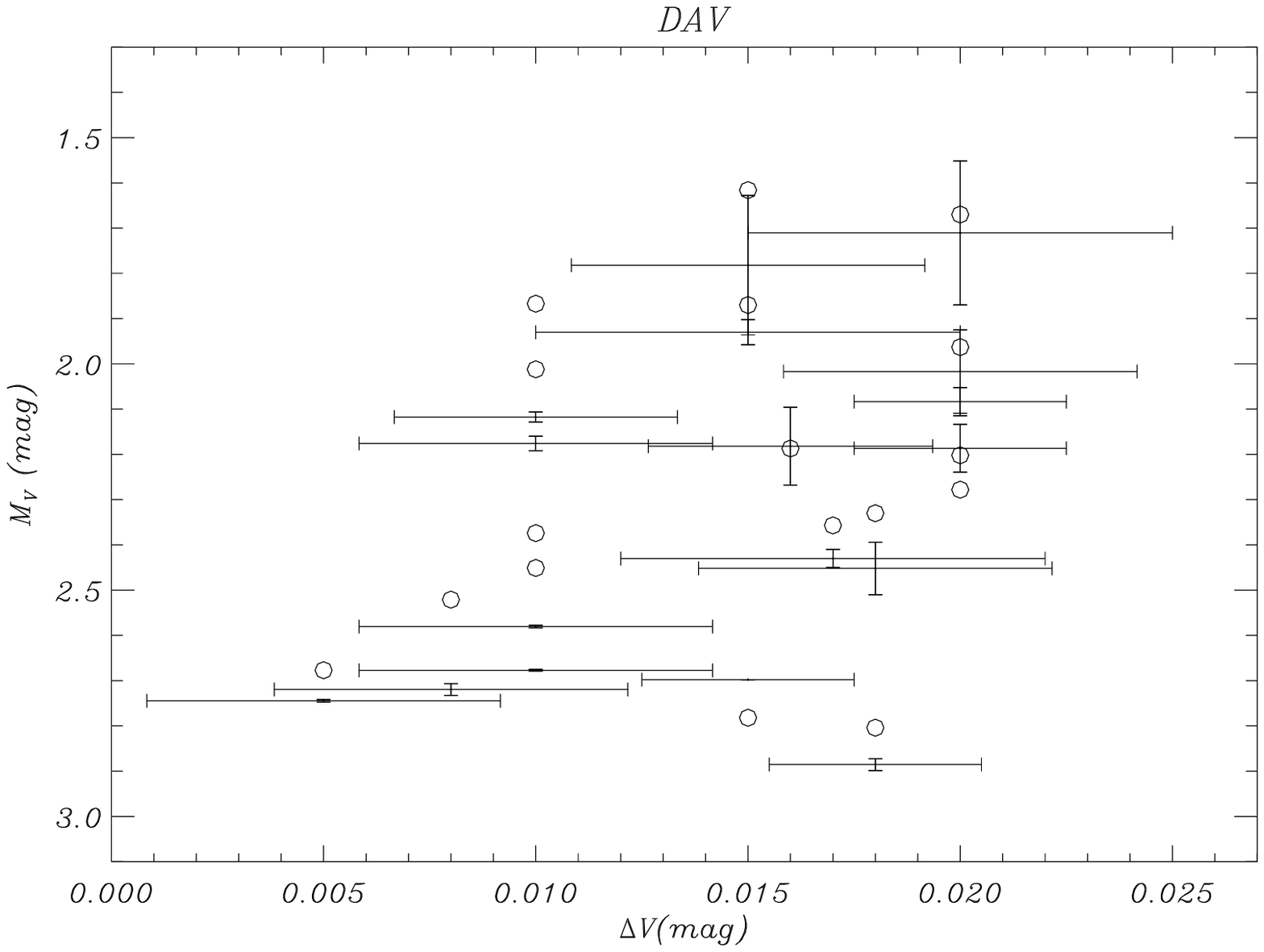}
\includegraphics[width=9cm]{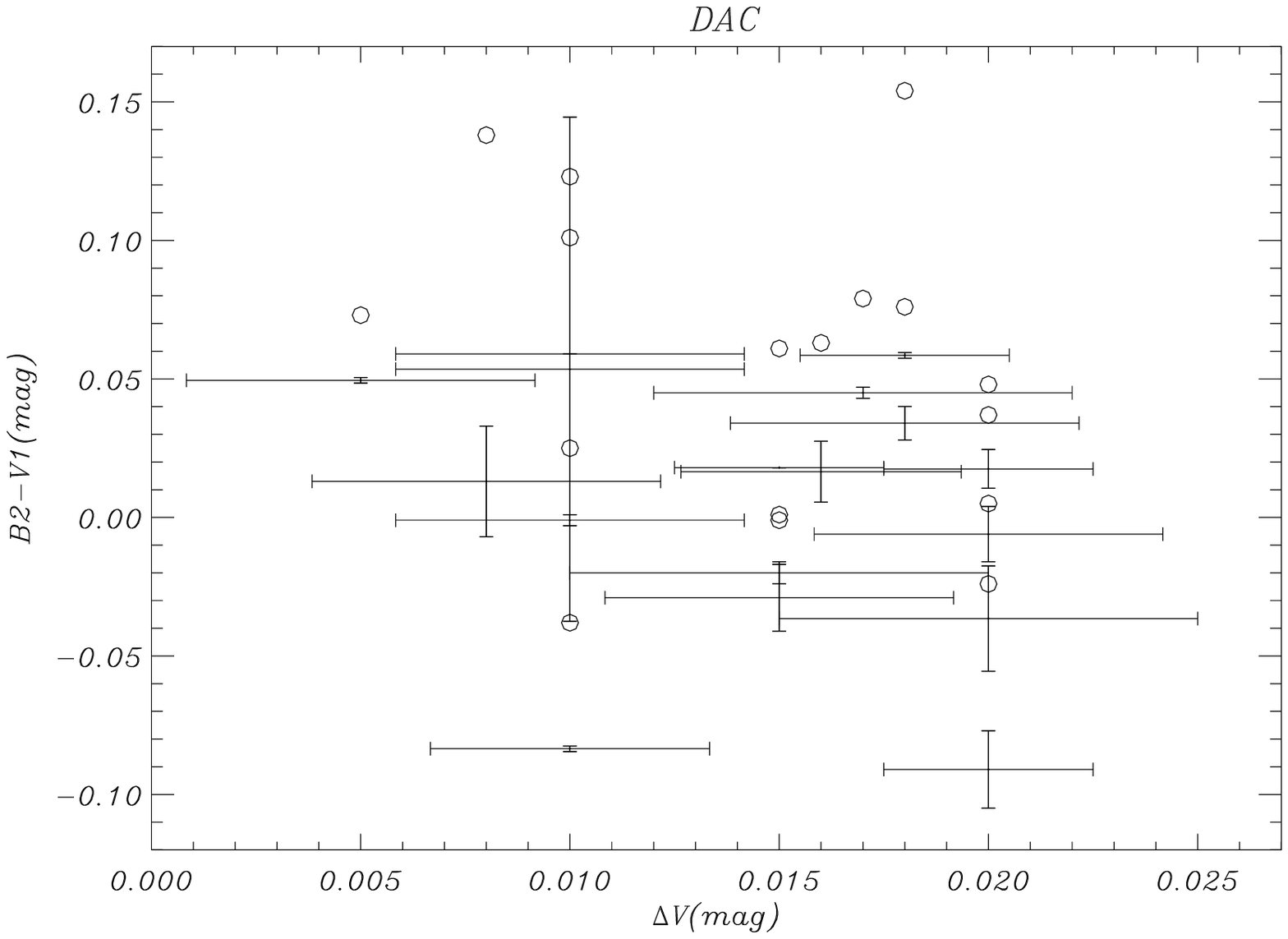}
\includegraphics[width=9cm]{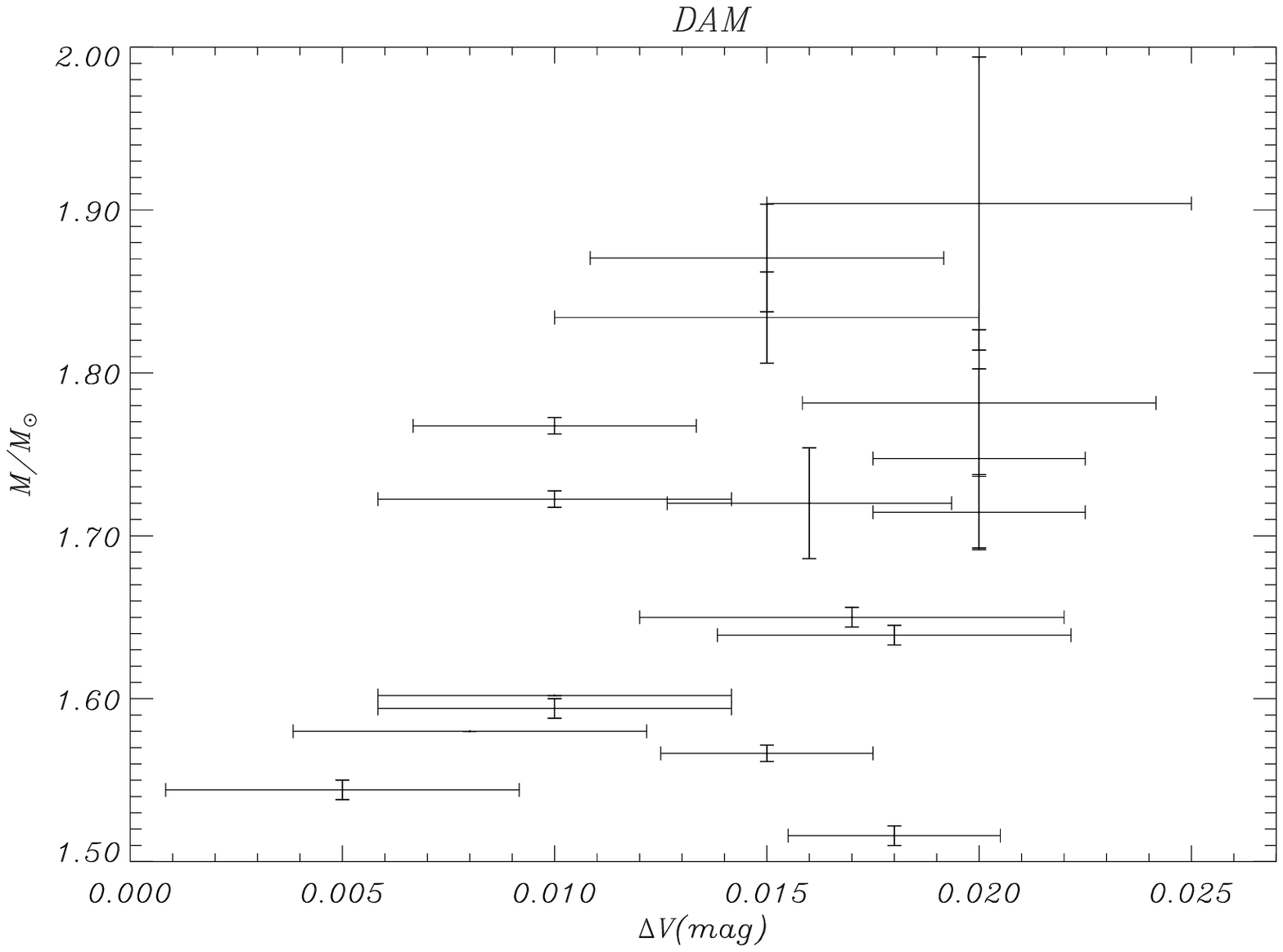}
\includegraphics[width=9cm]{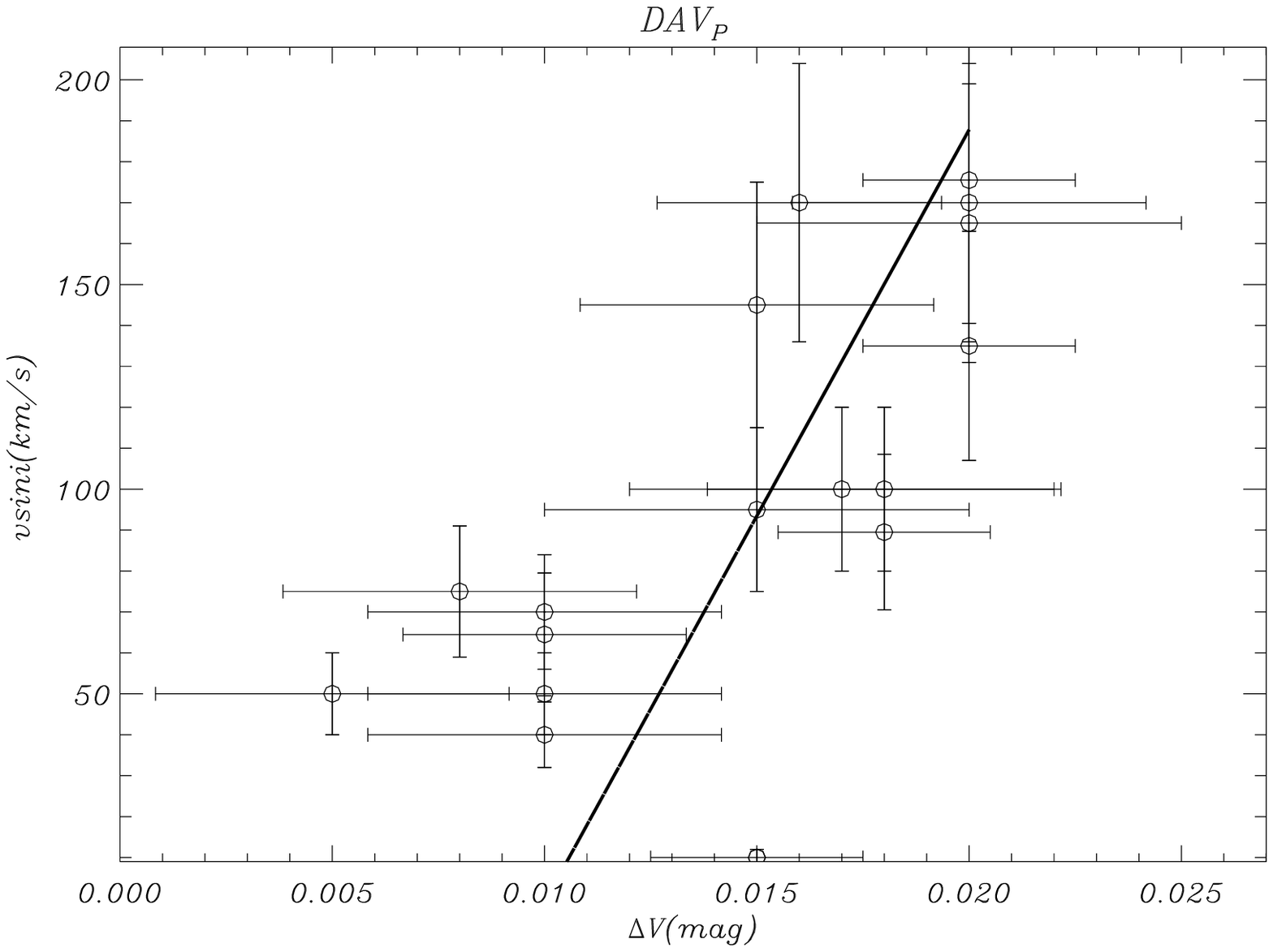}
\includegraphics[width=9cm]{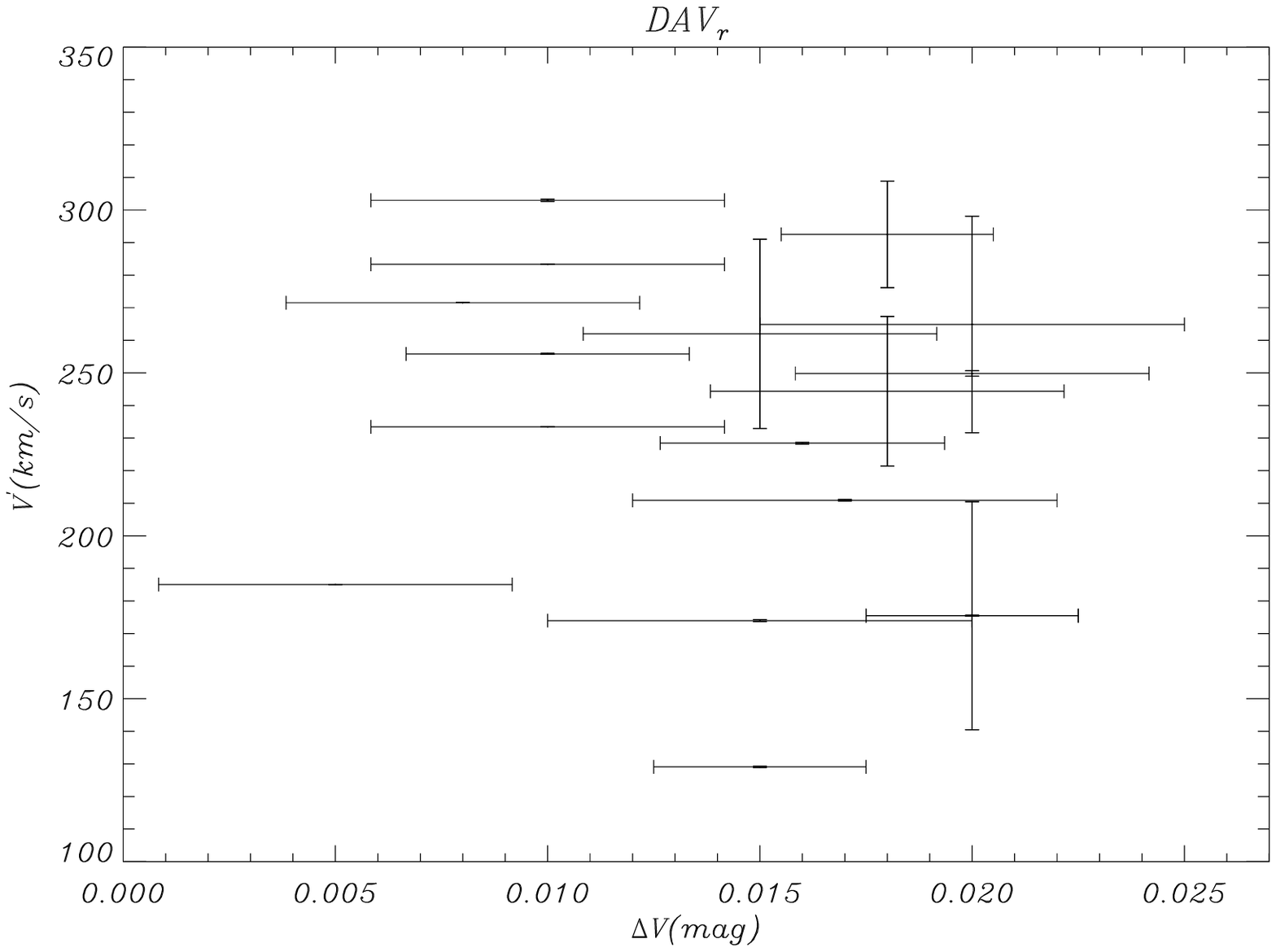}
\includegraphics[width=9cm]{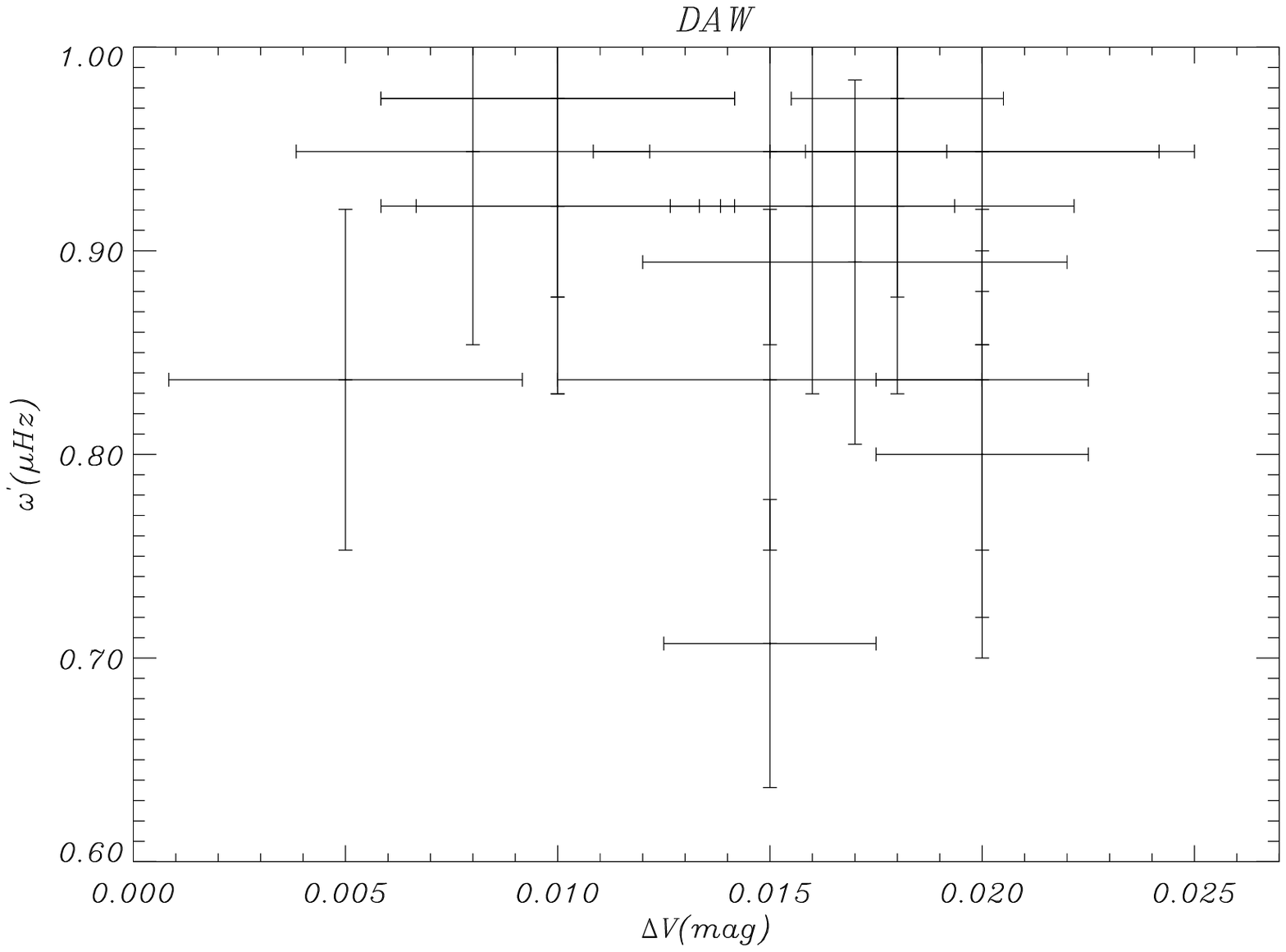}
\includegraphics[width=9cm]{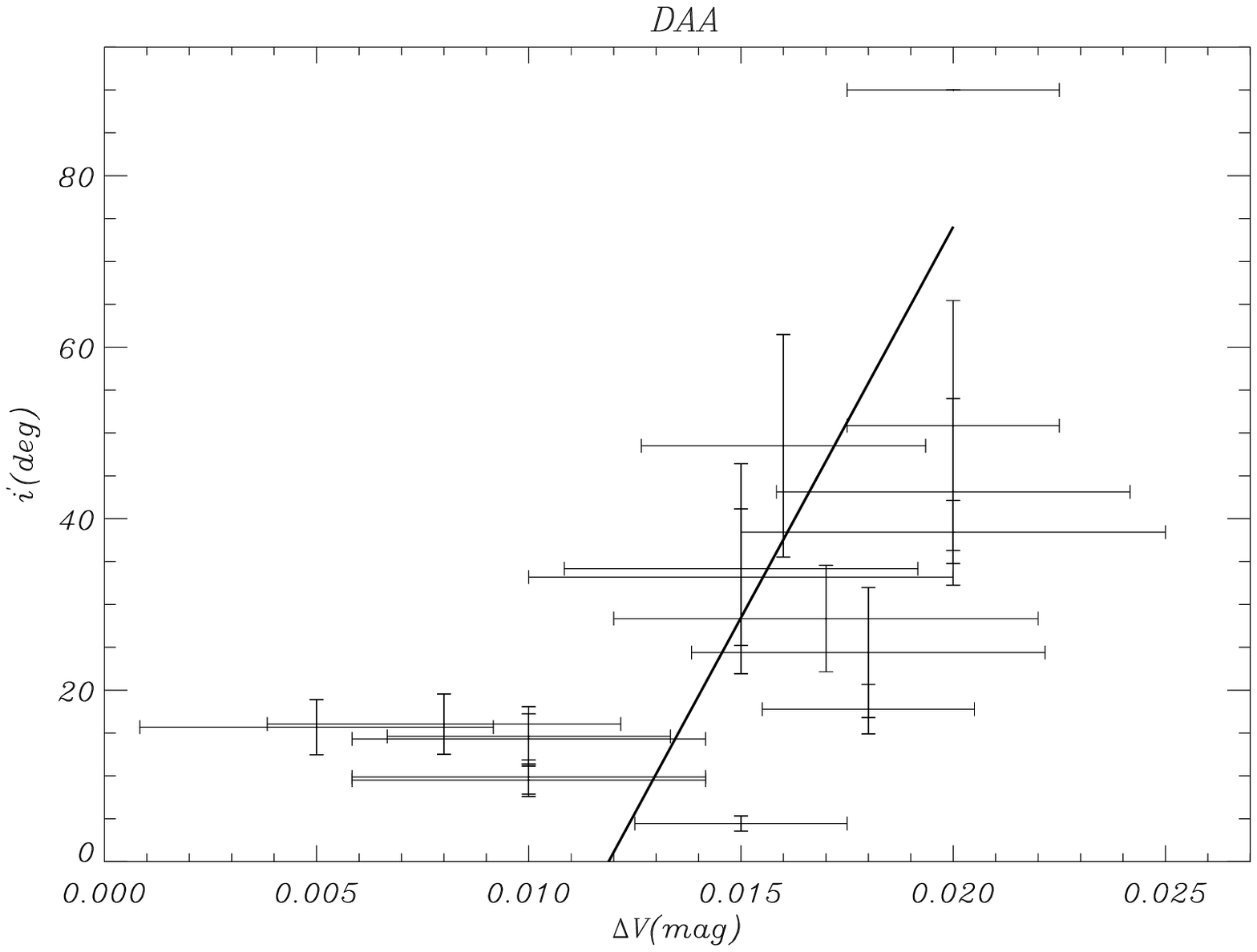}
\caption{Correlation diagrams for a set of photometric parameters
obtained from the correction for rotation (see
Table~\ref{tab-corr}). Circles represent the  \dss before
correcting for the effect of rotation. Crosses represent the error
bars (See Section~4 for details).} \label{fig-stat}
\end{figure*}

For corrected parameters
in Table~\ref{tab-corr}, the error bars correspond to values
obtained using $\pm10 \%$ in the observed $\vsini$. In
Fig.~\ref{fig-stat} we present the different diagrams with the
parameters corrected for rotation versus $\aosc$, with the
error bars for both.

\subsection{The correlation technique}

We computed the Spearman rank-order correlation coefficient,
$\rs$. This coefficient is preferred to the classic Pearson one
(also computed as a complement) as it is more robust to outliers
and does not presuppose a linear relation. For the error made in
the correlation coefficient, we use the following expression from
standard statistics,
\ \\
\begin{center}
   $\sigma(\rs)^2=\dst\frac{(1-\rs)^2}{n-1}\,
   \left(1+\frac{11\,\rs^2\,}{2n}\right)+\o(n^{-3})$
\end{center}
\ \\
where n is the dimension of the sample. In addition to this, we
computed, for the Spearman's coefficient, the parameter p which gives the
probability of deviation from the null-hypothesis for this
coefficient.

On the other hand, for the $\rs$ values indicating a reasonable
degree of correlation, we also search for a possible functional
dependence. In such cases, we have used an orthogonal distance
regression routine \citep[see]{Press89} which combines data sets
with errors in both variables.

\subsection{Results and discussion}

We have applied the statistical methods described hereabove to all
parameters corrected for rotation computed in Section~4. In
Table~\ref{tab-stat} are listed the coefficients obtained for
correlations between these parameters and the $\aosc$.

We now analyze the results obtained for the absolute
magnitude and the colour index. These two photometric parameters
have been generally used in the literature for statistical works
on \dss. Particularly, in a multi-correlate analysis made by
Antonello et al.~(1981), a significant individual correlation
between $\aosc$ and $\mv$ is found.
 No strong correlation is found here neither for $\mv$ nor for
 $\bv$, however we notice that the correlation coefficient is
approximately two times better for corrected values of $\mv$ than
for non-corrected ones ($\rs=0.26$). They are of the same order in
the case of $\bv$ ($\rs\sim0.25$). We thus conclude, following
\citep{Sua01}, that it is worthwhile to take into
account such a correction in this kind of study. We attribute the
low value obtained for the  coefficient in the case of the
magnitude to the fact that our sample is still too limited.

A less classical parameter for such studies is the projected
rotational velocity, $\vsini$, for which an unexpected significant
correlation is found ($\rs=0.76$). This parameter has been
considered before in \citet{LiMi99}, however no correlation was
calculated. This striking result has two possible explanations: a
physical effect (linked to $\vrp$ or maybe rather $\op$) and/or a
geometric effect (associated with \ip). Making use of the
different stellar parameters we have access to, we tried to go
further to understand the origin of this correlation. We can
notice that there is no significant correlation neither for the
$\vrp$ parameter nor for the $\op$ parameter.

On the other hand, for \ip, another significant correlation is
found ($\rs=0.75$) such that, the higher is the value of $\aosc$
the larger is \ip. These results suggest that the correlation
observed for $\vsini$ has a pure geometric origin. At this point,
we tried to figure out if a possible bias could be induced by the
method used to correct parameters for rotation. We can only figure
out a possible influence of our limited sample of stars. In
addition, the correlation found between $\vsini$ and $\aosc$ does
not involve the correction method.

\begin{table}
  \begin{center}
    \caption{Linear functional dependence between $\vsini$, \ip and
    $\aosc$. The linear fit parameters are $a$ ($\times10^3$) and $b$ following
    $y=a\aosc+b$. Their corresponding
    errors are $\sigma(a)$ ($\times10^3$) and $\sigma(b)$. $\chi^2$ represents the minimum Chi-Square of the linear fit and $q$
    is a scalar between 0 and 1 giving the probability that a \emph{right}
    model would give a value equal or larger than the observed $\chi^2$.}
    \vspace{1em}
    \renewcommand{\arraystretch}{1.2}
    \begin{tabular}[h]{ccccccc}
      \hline
        y & $a$ &  $\sigma(a)$ & ($b$) & $\sigma(b)$  & $\chi^2$ & $q$\\
      \hline
      \ \\
     $\vsini$ & 1.88 & 1.12 & -189.64 & 305.430 & 3.997 & 0.997\\
     \ip      & 9.12 & 5.5  & -108.3  & 156.69 & 4.13 & 0.997 \\

   \hline
      \end{tabular}
    \label{tab-linfit}
  \end{center}
\end{table}

In the framework of the classical description of the oscillations,
which attributes to each mode a given spherical harmonic, the
preceding result would first confirm that the variation of
flux observed in \ds light curves are not dominated by the radial
modes, which visibility coefficients are independent of \ip. This
is in agreement with the results obtained by \citet{Hernandez98}
 who showed that they could not find more than two
radial modes per star, in oscillation spectra, for a group of
selected stars in the Praesepe cluster.

The $\aosc$ value being considered as the maximum constructive
interference of the observed oscillation modes, the present
results would suggest that, at least, some modes with visibility
coefficients increasing with \ip are favored in the \dss
oscillations compared with other modes.

As can be seen in \citet{Pesnell85}, for instance, modes
corresponding to this description are the \emph{sectoral} modes
($m=\pm\ell$), in opposition to the \emph{zonal} modes
($(\ell=1,m=0)$,$(\ell=2,m=0)$) which visibility coefficient is
low for $i\rightarrow0$, and $(\ell=2,m=\pm1)$ which visibility
coefficient is low for $i\rightarrow0$ and for $i\rightarrow90$.
In this context, our results would suggest that the \ds
oscillations are dominated, at least to some extent, by modes with
$m=\pm\ell$. If confirmed, this interpretation would constitute an
important new element in the problem of mode identification of
\dss.

However, we want to stress that the classical spherical harmonics
system might not be the best context to interpret this result. It
is well known now, that \dss are fast rotators (this is confirmed
here where, for our sample of \dss, large $\vr$ values are found)
and that the effect of rotation
affects their structure to a point where the description in terms
of spherical harmonics might become questionable. An alternative
representation is being developed by Lignieres~\citep{Lig01} by
means of a non-perturbative numerical method. Their results obtained for homogeneous
ellipsoids show modes with geometric horizontal features very
different from classical spherical harmonics. These results are
preliminary and have to be applied to models closer to stellar
internal structure, but they might be indicative of relevant
geometric effects.

For projected velocity and \ip, we have also searched for a
functional dependence. In Table~\ref{tab-linfit} their linear fit
parameters are listed. We have to keep in mind that the
coefficients of these linear fits might be sensitive to eventual
systematic errors, which were neglected when searching for
correlations (see Section~5). While good confidence values ($q$)
are obtained, we can appreciate from values of $\chi^2$, that
scatter is important as well as errors in fit coefficients. This
situation can be improved by increasing the number of target
stars.

Finally, no correlation is found for the mass.

\section{Conclusions}
For a sample of 17 \dss, we report here the results of a search
for correlations between the oscillation amplitude and a set of
estimated fundamental parameters.

In this process we confirm the interest of applying correction for
fast rotation, as suggested by \citep{Sua01} for
this kind of study.

We found a significant correlation coefficient for the projected
velocity ($\vsini$) and show that it is due to the strong
correlation between the oscillation amplitude and an estimation
of the angle of
inclination of the star, \ip. We observe that the oscillation
amplitudes increase with \ip.
 This striking result is discussed in terms of the visibility
of modes. In the context of spherical harmonics, this suggests
that the oscillations of \dss would follow a selection rule in
favor of \emph{sectoral} modes (i.e. $m=\pm\ell$ modes).

An alternative interpretation might turn up with future
developments of the work initiated by Lignieres~\citep{Lig01}.

A future increase of our sample of \dss in clusters would help to
refine the present results.

\begin{table*}
  \begin{center}
    \caption{Photometric data for the 17 \dss of the 5 open clusters considered. Different columns
    represent the star HD (SAO) number, the absolute magnitude, the $\bv$ colour index, the oscillation
    amplitude $\aosc$, the projected velocity $\vsini$,the spectral type and the class of luminosity, the time
    observation weight (see Section~2) and the cluster they belong to.}
    \vspace{1em}
    \renewcommand{\arraystretch}{1.2}
    \begin{tabular}[h]{ccccccccc}
      \hline
        Star & $\mv$  & $\bv$ & $\aosc$ &  $\vsini$ & Sp & Weight & Cluster \\
        & (mag)  & (mag) &  (mag) & $(\mathrm{km\,s^{-1}})$ \\
      \hline
      \object{HD\,74050} & 1.62 & 0.001  & 0.015 & 145 & A7V   & 2 & Praesepe\\
      \object{HD\,74028} & 1.67 & 0.005  & 0.020 & 165 & A7III & 1 & Praesepe\\
      \object{HD\,73345} & 1.87 & -0.001 & 0.015 & 95  & F0V   & 1 & Praesepe\\
      \object{HD\,73175} & 1.96 & 0.037  & 0.020 & 170 & A9V   & 2 & Praesepe\\
      \object{HD\,73798} & 2.12 & 0.063  & 0.016 & 170 & F0V   & 4 & Praesepe\\
      \object{HD\,73450} & 2.10 & 0.048  & 0.020 & 135 & A7V   & 3 & Praesepe\\
      \object{HD\,73746} & 2.36 & 0.079  & 0.017 & 110 & A9V   & 1 & Praesepe\\
      \object{HD\,23156} & 2.01 & -0.038 & 0.010 & 65  & A7V   & 3 & Pleiades\\
      \object{HD\,23567} & 2.80 & 0.154  & 0.018 & 90  & A9V   & 4 & Pleiades\\
      \object{HD\,23607} & 2.78 & 0.061  & 0.015 & 10  & A7V   & 4 & Pleiades\\
      \object{HD\,23643} & 2.28 & -0.024 & 0.020 & 175 & A3V   & 4 & Pleiades\\
      \object{HD\,27397} & 2.33 & 0.076  & 0.018 & 100 & F0V   & 2 & Hyades\\
      \object{HD\,27459} & 1.87 & 0.025  & 0.010 & 70  & A9V   & 2 & Hyades\\
      \object{HD\,27628} & 2.45 & 0.101 & 0.010 & 40  & Am    & 2 & Hyades\\
      \object{HD\,107513}& 2.68 & 0.073  & 0.005 & 50  & Am    & 2 & Coma Ber\\
      \object{SAO\,38754} & 2.52 & 0.138  & 0.008 & 75  & F0IV       & 2 & \aper\\
      \object{SAO\,38788} & 2.37 & 0.123  & 0.010 & 50  & A8V      & 2 & \aper\\
      \hline
      \end{tabular}
    \label{tab-phot}
  \end{center}
\end{table*}

\begin{table*}
  \begin{center}
    \caption{Non-rotating associate models for the \dss in
the 5 open clusters considered. The different columns represent the
range of values obtained from the correction for the effect of
rotation (see section\,4) for: $\mv$ the absolute magnitude; $\bv$
the colour index; $M$ the mass (in $M_{\sun}$); $R$ the radius (in
$R_{\sun}$); $\teff$ the effective temperature (log, in K); $L$
the luminosity (in $L_{\sun}$); $\op$ the rotation rate;
$\nu_{\mathrm{rot}^{\prime}}$ the rotation frequency ($\mu\mbox{Hz}$); \ip the angle of
inclination (deg), and finally $g$ the gravity (log, in
$\mbox{cm\,s}^2$).}\vspace{1em}
    \renewcommand{\arraystretch}{1.2}
    \begin{tabular}{ccccccccccc}
      \hline
       Star & $\mv$  & $\bv$ & $\mmsun$ & $R/R_{\sun}$ & $log\,\teff$ \\
            & $L/L_{\sun}$ & $\op$  & $\nu_{\mathrm{rot}^{\prime}}$ & \ip & $log\,g$ \\
      \hline

      \object{HD\,74050} & 1.705--1.859  & -0.035--(-0.023) & 1.89--1.86 & 2.09--2.04  & 3.90--3.89 \\
                         & 15.65--14.46& 0.90--0.90 & 25.08--25.82& 40.28--28.04  & 4.07--4.08\\
      \object{HD\,74028} & 1.631--1.790  & -0.046--(-0.027) & 1.95--1.86 & 2.19--2.05  & 3.90--3.89 \\
                         & 18.06--14.65& 0.90--0.90 & 23.64--25.68& 40.28--36.59  & 4.04--4.08\\
      \object{HD\,73345} & 1.916--1.944  & -0.022--(-0.018) & 1.85--1.82 & 2.03--1.99  & 3.89--3.89 \\
                         & 14.27--13.53& 0.70--0.70 & 25.77--26.19& 37.16--29.21  & 4.01--4.09\\
      \object{HD\,73175} & 1.971--2.063  & -0.011--(-0.001) & 1.80--1.76 & 1.97--1.91  & 3.89--3.88 \\
                         & 12.83--11.52& 0.90--0.90 & 26.88--27.83& 48.57--37.69  & 4.11--4.12\\
      \object{HD\,73798} & 2.139--2.225  &  0.011--0.022    & 1.74--1.70 & 1.88--1.84  & 3.88--3.88 \\
                         & 10.91--10.05& 0.85--0.85 & 28.25--28.91& 54.98--42.02  & 4.13--4.14\\
      \object{HD\,73450} & 2.160--2.213  &  0.014--0.021    & 1.73--1.70 & 1.86--1.84  & 3.88--3.88 \\
                         & 10.62--10.05& 0.70--0.70 & 28.35--28.78& 58.15--43.56  & 4.13--4.14\\
      \object{HD\,73746} & 2.420--2.440  &  0.044--0.046    & 1.65--1.65 & 1.75--1.74  & 3.87--3.87 \\
                         & 8.48--8.32  & 0.80--0.80 & 30.51--30.69& 31.45--25.24  & 4.17--4.17\\
      \object{HD\,23156} & 2.112--2.123  & -0.084--(-0.083) & 1.77--1.76 & 1.52--1.52  & 3.93--3.94 \\
                         & 11.86--11.72& 0.85--0.85 & 39.09--39.15& 16.34--12.89  & 4.32--4.32\\
      \object{HD\,23567} & 2.879--2.892  &  0.059--0.058    & 1.52--1.51 & 1.41--1.41  & 3.88--3.87 \\
                         & 5.57--5.46  & 0.95--0.95 & 40.63--40.64& 19.22--16.34  & 4.32--4.31\\
      \object{HD\,23607} & 2.698--2.698  &  0.018--0.018    & 1.57--1.56 & 1.43--1.43  & 3.89--3.88 \\
                         & 6.63--6.50  & 0.50--0.50 & 39.87--39.93&  4.88--4.00    & 4.32--4.32\\
      \object{HD\,23643} & 2.068--2.099 & -0.098--(-0.086) & 1.77--1.72 & 1.523--1.49  & 3.94--3.93 \\
                         & 12.00--10.63& 0.69--0.59 & 38.89--38.25& 90.00--90.00  & 4.32--4.32\\
      \object{HD\,27397} & 2.423--2.481  &  0.037--0.031    & 1.64--1.64 & 2.029--2.02  & 3.87--3.87 \\
                         & 8.27--8.11  & 0.85--0.85 & 24.44--24.48& 28.18--20.59  & 4.19--4.19\\
      \object{HD\,27459} & 2.168--2.184  & -0.002--0.000    & 1.72--1.72 & 2.152--2.15  & 3.89--3.88 \\
                         & 10.55--10.68& 0.95--0.95 & 22.98--22.96& 15.77--12.84  & 4.16--4.16\\
      \object{HD\,27628} & 2.579--2.582  &  0.059--0.059    & 1.60--1.60 & 2.072--2.07  & 3.87--3.87 \\
                         & 7.20--7.21  & 0.85--0.85 & 23.39--23.39& 10.86--8.87   & 4.21--4.20\\
      \object{HD\,107513}& 2.743--2.746  &  0.050--0.049    & 1.55--1.54 & 1.499--1.75  & 3.87--3.87 \\
                         & 6.30--6.17  & 0.70--0.70 & 37.18-29.44& 17.29--14.07  & 4.27--4.27\\
      \object{SAO\,38754}& 2.713--2.726  &  0.023--0.003    & 1.58--1.580 & 1.455--1.45  & 3.88--3.88 \\
                         & 6.56--6.55  & 0.90--0.90 & 39.54-39.58& 17.79--14.28  & 4.31--4.31\\
      \object{SAO\,38788}& 2.678--2.676  &  0.099--0.008    & 1.59--1.591 & 1.461--1.45 & 3.89--3.88 \\
                         & 6.93--6.80  & 0.95--0.95 & 39.58--39.79& 10.45--08.55  & 4.31--4.31\\
      \hline
      \end{tabular}
    \label{tab-corr}
  \end{center}
\end{table*}

\bibliography{/mercure/suarez/LATEX/REFERENCES/ref-generale}
\bibliographystyle{aa}

\end{document}